\documentclass[12pt,a4paper,openright]{article}
\usepackage{graphicx}
\usepackage[english]{babel}
\usepackage[T1]{fontenc}
\usepackage[latin2]{inputenc}
\usepackage{amsmath}
\usepackage{amssymb}
\begin{document}


\newcommand{\be}{\begin{equation}}
\newcommand{\ee}{\end{equation}}
\newcommand{\ba}{\begin{eqnarray}}
\newcommand{\ea}{\end{eqnarray}}
\newcommand{\bt}{\begin{tabular}}
\newcommand{\et}{\end{tabular}}
\newcommand{\bc}{\begin{center}}
\newcommand{\ec}{\end{center}}
\newcommand{\ben}{\begin{enumerate}}
\newcommand{\een}{\end{enumerate}}
\newcommand{\bi}{\begin{itemize}}
\newcommand{\ei}{\end{itemize}}
\newcommand{\bmpage}{\begin{minipage}}
\newcommand{\empage}{\end{minipage}}
\newcommand{\disty}{\displaystyle}

\def \al{\alpha}
\def\bet{\beta}
\def \df {{\rm d}}
\def \del{\delta}
\def \dg{\dagger}
\def\eps{\epsilon}
\def\veps{\varepsilon}
\def\ga{\gamma}
\def\Ga{\Gamma}
\def \si {\sigma}
\def\la {\lambda}
\def\latl{\tilde{\lambda}}
\def\La {\Lambda}
\def\om {\omega}
\def\Om {\Omega}
\def\K {\tilde{K}}
\def\J {\tilde{J}}
\def\Igr{ {\bf I} }
\def\kgr{\bf k}
\def\Pgr{\bf P}
\def\Ngr{\bf N}
\def\Qgr{\bf Q}
\def\Rgr{\bf R}
\def\Sgr{ {\bf S } }
\def\Zgr{\bf Z}
\def\bul{$\bullet\;$}
\def\bulbul{$\bullet\bullet\;$}
\def\Ccal{ {\cal C}}
\def\Bcal{ {\cal B}}
\def\Scal{ {\cal S}}
\def\Pcal{ {\cal P}}
\def\Rcal{ {\cal R}}
\def\Ecal{ {\cal E}}
\def\Mcal{ {\cal M}}
\def\Ocal{ {\cal O}}
\def\Gcal{ {\cal G} }
\def\Hcal{ {\cal H} }

\def\imply{\;\Longrightarrow\;}
\def\sgn{{\rm sgn} }


\def\six{ \frac{1}{2}\left[ 
\begin{array}{cc}
0&1\\
1&0
\end{array}
\right]}

\def\siy{ \frac{1}{2}\left[ 
\begin{array}{cc}
0&-i\\
i&0
\end{array}
\right]}

\def\siz{ \frac{1}{2}\left[ 
\begin{array}{cc}
1&0\\
0&-1
\end{array}
\right]}

\def\siv{\vec{\si}}
\def\spinup{\left[ 
\begin{array}{c}
1\\
0
\end{array}
\right]}

\def\spindown{\left[ 
\begin{array}{c}
0\\
1
\end{array}
\right]}

\def\siisij{\vec{\si}_i\cdot\vec{\si}_j}

\def\kket{\rangle}
\def\brak{\langle}

\def\pol{\frac{1}{2}}
\def\sq2i{\frac{1}{\sqrt{2}}}

\def\bscr{\begin{scriptsize}}
\def\escr{\end{scriptsize}}
\def\plak{ {\Box} }
\def\sngl{|{\sf singlet}\kket}
\def\tryp{|{\sf tryplet}\kket}
\def\quin{|{\sf quintet}\kket}

\def\pppp
{\bscr\left|
\begin{array}{cc}
+&+\\
+&+
\end{array}
\right\rangle
\escr
}

\def\mppp
{\bscr\left|
\begin{array}{cc}
-&+\\
+&+
\end{array}
\right\rangle
\escr
}
\def\pmpp
{\bscr\left|
\begin{array}{cc}
+&-\\
+&+
\end{array}
\right\rangle
\escr
}

\def\ppmp
{\bscr\left|
\begin{array}{cc}
+&+\\
-&+
\end{array}
\right\rangle
\escr
}
\def\pppm
{\bscr\left|
\begin{array}{cc}
+&+\\
+&-
\end{array}
\right\rangle
\escr
}

\def\ppmm
{\bscr\left|
\begin{array}{cc}
+&+\\
-&-
\end{array}
\right\rangle
\escr
}
\def\mmpp
{\bscr\left|
\begin{array}{cc}
-&-\\
+&+
\end{array}
\right\rangle
\escr
}
\def\pmmp
{\bscr\left|
\begin{array}{cc}
+&-\\
+&-
\end{array}
\right\rangle
\escr
}
\def\mppm
{\bscr\left|
\begin{array}{cc}
-&+\\
-&+
\end{array}
\right\rangle
\escr
}
\def\pmpm
{\bscr\left|
\begin{array}{cc}
+&-\\
-&+
\end{array}
\right\rangle
\escr
}
\def\mpmp
{\bscr\left|
\begin{array}{cc}
-&+\\
+&-
\end{array}
\right\rangle
\escr
}

\title{Quantum Monte Carlo scheme for frustrated Heisenberg antiferromagnets} 
\author{Jacek Wojtkiewicz\\
{\em Department for Mathematical Methods in Physics}\\
{\em Faculty of Physics, Warsaw University}\\
{\em Ho\.{z}a 74, 00-682 Warszawa, Poland}\\
e-mail: wjacek@fuw.edu.pl
}
\maketitle
\abstract{
When one tries to simulate quantum spin systems by the Monte Carlo method, often
the 'minus-sign problem' is encountered. In such a case, an application of
probabilistic methods  is not possible.
 In this
 paper the method has been proposed how to avoid the minus sign problem for
 certain class of frustrated Heisenberg models. The systems where this method
 is applicable are, for
 instance, the pyrochlore lattice and the $J_1-J_2$ Heisenberg
 model. The method works in singlet sector. It relies on expression of wave
 functions in dimer (pseudo)basis and writing down  the Hamiltonian as a sum
 over plaquettes. In such a formulation, matrix elements of the exponent of
 Hamiltonian are positive.
}

\section{Introduction}

The systems considered in this paper are Heisenberg models for spin one-half,
with competing antiferromagnetic (AF) interactions  -- i.e. {\em frustrated}
 ones:
\[
\displaystyle
H_\La={\sum}_{i,j\in\La}J_{ij}\, {\bf s}_i\cdot {\bf s}_j,
\]
where ${\bf s }$ is a vector of Pauli matrices, $\La\subset\mathbb{Z}^d$, 
$J_{ij}>0$ (AF case).

Frustrated systems are very interesting and hard to analyse and
understand, both in classical version and especially in the quantum
case. 
The source of these difficulties traces back to the large
ground-state degeneracy in the classical version. The prototype of such system
is AF Ising model on triangular lattice \cite{WannHout}. Such systems are
very sensitive to perturbations. A consequence  is a possibility of
very complicated ground-state and finite temperature phase diagrams. This situation
takes place, for
instance, in an ANNNI model \cite{FishSzp} (infinite number of phases,
devil's staircase, etc). Besides of numerous efforts
and important results \cite{FishSzp}, \cite{FishSel},  \cite{DinMaz} 
 (for reviews, see \cite{Selke}, \cite{FrIs}, \cite{FrSS}),
 full treatment of such systems is not worked out so
 far.

 The situation for quantum frustrated antiferromagnet is even less
clear.  
It is generally suspected that -- in the case of strongly frustrated
systems -- the ground state emerging as a
linear combination of many classical configurations is a
featureless, ``spin liquid'' state, i.e. state without
 long-range ordering, where 
 correlation  functions fall-off exponentially
 \cite{Anderson73}. However, one cannot 
 exclude another scenario: ``order by disorder'' -- exotic orderings
 absent in a classical version of these models. Such scenarios are
 moreover sensitive to the underlying lattice structure. (For a
 review, see for instance
 \cite{Moessner}). To my best knowledge, no general definite
 conclusions have been obtained so far.

Among frustrated lattices, perhaps the most popular ones are:
triangular;  kagom\'{e};
pyrochlore;  square lattice with 'crossing bonds' (called also the $J_1 - J_2$
model). This last case is particularly interesting due to its possible relation
with high-temperature superconductivity: Quite often one considers the  $t-t'$
two-dimensional  Hubbard model as a 'minimal model' for HTSC
\cite{tt1HuM}. Behaviour of this last model is still not fully understood. A
natural starting point in such a study  is the limiting case:  half-filling and
large coupling constant; under these conditions, the $t-t'$ Hubbard model
simplifies to $J_1-J_2$ Heisenberg model.
 
As a sample of natural questions  in the study of frustrated systems one
can mention the  following ones   
(for definiteness, let us concentrate on the  $J_1-J_2$ Heisenberg model):
\bi
\item Nature of ground state: For which range of  values of the ratio
  $\alpha=J_2\slash{}J_1$ we have an antiferromagnetic (Neel) ordering? Is
  there a spin-liquid  state for strong frustration?
\item Describe the nature of crossover between ordered and disordered state
upon increasing frustration.
\ei

Exact results on the area of frustrated models are rather rare.  For
particular forms of interactions, 
there exist exact results for ground states, obtained by AKLT  \cite{AKLT} as
well as related results  
\cite{ZittKlumpEtAl}. It is however unclear if they can be generalized to more
general forms of interactions. Some general properties of frustrated systems
have been obtained in \cite{LS1}, \cite{LS2}, \cite{S3}
 (they are important in the context of this paper). One can mention also quite
 a few approximate reliable results, for instance \cite{SorellaEtAl} (based on
 BCS-like 
 ansatz on wave function).


One of general tools used to calculate the partition function $Z_\La$ and
thermodynamic functions for quantum spin systems is an
 application of {\em Lie--Trotter product formula}. Let us describe the
 general setup
 of certain version thereof, i.e. {\em the  Suzuki approach} \cite{Suzuki}.

Lie--Trotter product formula  \cite{Suzuki}, \cite{ReedSimon1} states
that if $A,B$ -- finite-dimensional matrices, then

\[
e^{A+B} = \lim_{n\to\infty}\left( e^{\frac{A}{n}} e^{\frac{B}{n}}
\right )^n. 
\]

Using this formula, one calculates $Z_\La$ in the following way:

\ben
\item Write:
\[
H=H_1+H_2,
\]
in such a way that $H_1$, $H_2$ are sums of {\em commuting} operators.
\item Using Lie--Trotter formula, we have:
\[
Z = {\rm Tr}e^{-\bet H} = {\rm Tr}e^{-\bet (H_1+H_2)}=
\]
\[
={\rm Tr}
\left[
 \lim_{n\to\infty}\left( e^{\frac{-\bet H_1}{n}} e^{\frac{-\bet
 H_2}{n}} \right)^n
\right] =
\]
\[
=\lim_{n\to\infty}\sum_{ \{\al_1\},\{\al_2\},\dots,\{\al_{2n}\}}
\brak \al_1 |  e^{\frac{-\bet H_1}{n}} |\al_2 \kket
\brak \al_2 |  e^{\frac{-\bet H_2}{n}} |\al_3 \kket
\dots
\]
\be
\cdot 
\brak \al_{2n-1} |  e^{\frac{-\bet H_1}{n}} |\al_{2n} \kket
\brak \al_{2n} |  e^{\frac{-\bet H_2}{n}} |\al_{1} \kket
\label{GenFormula}
\ee
(here $\{\al_i\}$ is a  basis in the Hilbert space of system states).
\een

If it happens that above matrix elements are {\em positive},
then the life is easier, as one can apply {\em probabilistic} techniques:
\bi
\item Monte Carlo method -- in numerical aspects; it is called the quantum Monte
Carlo \cite{deRaedt_qMC},
\item contour expansion techniques \cite{Kennedy85} or ``stochastic geometry''
    methodology in rigorous studies; as examples, one 
    can mention  spin chains \cite{AizenmanNachtergaele94}, or  
Bose-Hubbard models  \cite{ALSSY2004}, \cite{FeFroUel}.
\ei

For numerous important cases, matrix elements are positive. It is the case,
for instance, of the quantum Ising model in transverse magnetic field,
ferromagnetic Heisenberg model, XY model, Falicov-Kimball model (for a review,
see \cite{deRaedt_qMC}). These systems as well as numerous other ones have
been successfully studied with the use of quantum Monte Carlo method.

But alas! In general, matrix elements are {\em not} positive. (This is famous
``minus-sign problem'' in the quantum Monte Carlo method).

In certain cases, this problem can be overcomed. For instance, if one
considers antiferromagnetic Heisenberg model, then for simplest choice of
basis states in (\ref{GenFormula}) (Ising basis), the problem is present. But
it can be overcomed for the model on bipartite lattices, using more
sophisticated techniques  \cite{Manousakis}. This is also the case of the
Hubbard model on bipartite lattices and for half-filling
\cite{Hirsch}. Some results for frustrated antiferromagnets have been
reported \cite{Sandvik}. However
-- to my best knowledge -- the solution of the minus-sign problem is
still lacking  for general frustrated antiferromagnetic Heisenberg models.

{\em The goal of this paper is elaboration of the quantum Monte Carlo scheme
  for certain class of frustrated Heisenberg models. Using this scheme, matrix
  elements   obtained  are positive.}

This scheme concerns the $J_1-J_2$  Heisenberg model and holds under certain
conditions. There are:
\ben
\item Presence of reflection symmetry in the system,
 \item We restrict ourselves to the {\em  singlet sector} of the system
   (i.e. we assume that the total spin of the system is zero).
\een
The first assumption is not too restrictive and is technical only; it is
   necessary to apply the 
   Lieb-Schupp theorem (discussed below). The second one is more serious. But we
   can argue as follows. For certain class of frustrated antiferromagnetic
   Heisenberg    models (including the pyrochlore and $J_1-J_2$ model but not
   the triangular or kagom\'{e} lattices), we have Lieb and Schupp theorem
   \cite{LS1},\cite{LS2}, \cite{S3}   stating that the ground
   state of such systems is singlet. One then can hope that performing the MC
   simulation in the singlet sector  at finite temperature $T$, and then
   tending with $T$ to zero, we will obtain reliable properties of ground
   state of such a system. 

The sketch of the scheme is as follows. It is well known that positivity of
matrix elements is a problem of the choice of basis in the set of
'intermediate states' $|\al_i \kket
\brak \al_i |$ in formula (\ref{GenFormula}) (instructive examples can be
found in \cite{deRaedt_qMC}). If one chooses the  basis being a tensor product
of Ising states (it is perhaps the
simplest choice) as the basis of 'intermediate states', then some of matrix
elements of Hamiltonian are negative.

Assume however that we work in the singlet sector. It is known that every
singlet can be built up from {\em dimers}, i.e. two-spin wave functions of
total spin equal to zero. (This will be discussed in the 
Section~\ref{sec:GeneralSetup}). Such a form of singlets have been used in
numerous papers: \cite{Anderson73}, \cite{LiDoAn88}, \cite{Caspers},
\cite{Sutherland}. Consider now the system defined on the 
 $\mathbb{Z}^2$ lattice or, more generally, on a bipartite one (i.e. composed of
 two sublattices  called $A$ and $B$ in such a way that only neighbours of $A$
 type sites are $B$ type sites and vice versa). Moreover, let us impose the
 condition that one spin of every dimer belongs to $A$ sublattice and the
 second one to $B$ sublattice. Now, consider the model with nn interactions
 (or, more generally, with unfrustrated ones). It turns out that (for details
 see below Sec. \ref{sec:h2}) that matrix elements of the Hamiltonian are {\em
   non-negative}.  

But if we consider the  Heisenberg  model with  {\em
  frustrated} interactions (for definiteness, take the  $J_1-J_2$ Heisenberg
  model), then some of matrix elements are still {\em negative}.  How to cure this
  situation? The idea is as follows: Let us write the Hamiltonian as a sum of
  the {\em plaquette} 
  terms, i.e. four-spin Hamiltonians being defined on the $2\times 2$
  plaquettes on the lattice. One can hope that in such a situation, negative
  contributions coming from nnn interactions will be compensated by positive
  ones coming from nn interactions. 
{\em It turns out this is the case}: Matrix elements calculated with the use
  of plaquettes are {\em positive}. The calculations are presented in Sec.~\ref{sec:h4}. 

The paper is organized as follows. In the Sec.~\ref{sec:GeneralSetup}, the
general setup is introduced: construction of singlet wave functions from dimers
is explained, and the scalar product of two  singlet wave functions is
calculated and interpreted in geometrical terms. 
In Sec.~\ref{sec:h2}, the matrix elements of the Hamiltonian written as a sum
of two-spin interactions are calculated. 

The central part of the paper is Sec.~\ref{sec:h4}, where 
matrix elements of the Hamiltonian written as a sum
of four-spin (plaquette) interactions are calculated. Moreover, the (not quite
obvious) positivity of matrix elements is proved there. The
Sec.~\ref{sec:Conclus} contains 
summary and discussion of results obtained as well as considerations how
to generalize the results. In the Appendix, technical tools used in
calculations are presented (spectral resolution of self-adjoint operators
 and its application to two- and four-spin hamiltonians).

\section{Dimers, singlets and all that}
\label{sec:GeneralSetup}

Consider the {\em dimer}, i.e. the singlet wave function localized on sites $i,j$:
\[
(ij) = \frac{1}{\sqrt{2}} (|i_+ j_- \kket - |i_- j_+ \kket)
\]
Assume now that the total number of spins is even, i.e.  we are dealing with
 $2N$ spin system. Then every singlet wave function 
$\Psi_{2N}$ can be built from {\em dimers}: 
\be
\Psi_{2N}= \sum c_{i_1,\dots, i_N; j_1,\dots,j_N}
(i_1 j_1)(i_2 j_2) \dots (i_N j_N),
\label{ReprOfDimer}
\ee
\cite{LiDoAn88}, \cite{Caspers}, \cite{Sutherland}.
 This representation is non-unique for $N>1$ (the set of all such dimer products is an
 overcompleted set of vectors spanning all the space of singlets).

Consider now the {\em square}  lattice. It is bipartite one, and all
considerations below refer also to such lattices. Divide the lattice into
two kinds of sites: $A$- and $B$-type sites.
We demand that in (\ref{ReprOfDimer})
\be
i_k\in A,\;\;j_k\in B
\;\;\;\;\; \mbox{\rm for all} \;\;\; k=1,\dots,N.
\label{AuxiliaryCondition}
\ee
Also in this case the set of all  dimer products is an
 overcompleted set (for $N>2$) in the vector space of singlets.

\begin{figure}
\includegraphics{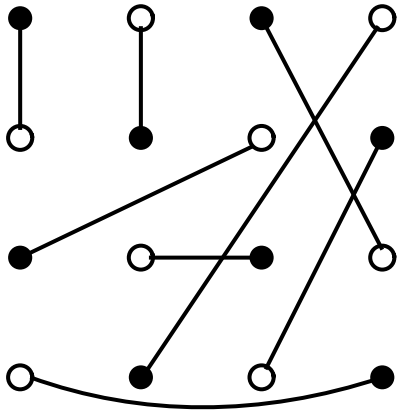}
\caption{An example of dimer wave function on $4\times 4$ lattice. Two sorts
  of sites of the bipartite lattice are represented by circles and heavy
  dots. Dimers are represented as 'bonds' linking lattice sites of 
  opposite kinds.}
\label{fig:fig1}
\end{figure}

Consider now some singlet wave function on the lattice, which is a
product of  dimers. Such a function possess a natural geometric
interpretation \cite{LiDoAn88}. Every dimer $(i_k j_k)$ can be illustrated as
a 'bond' linking lattice 
sites $i_k$ and $j_k$ (remember $i_k\in A$, $ j_k \in B$). 
Notice that every lattice site is occupied by the end of exactly one bond;
in the other  words, dimers are 'closely packed'. Such a situation is
illustrated on Fig.~\ref{fig:fig1}. 

\begin{figure}
\includegraphics{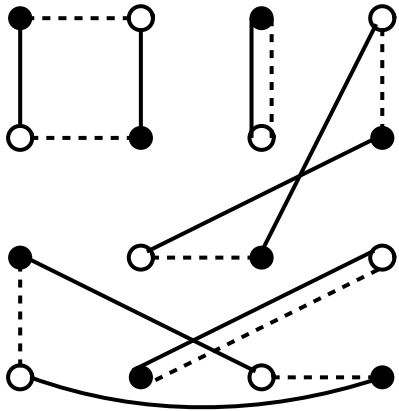}
\caption{Two dimer functions ($\Psi_1$: dashed line and $\Psi_2$: continuous
  line) and polygons formed by them. Every site is occupied by exactly one
   end of dimer  belonging to $\Psi_1$ and the same for $\Psi_2$. On the
  picture there are two trivial polygons (formed by 2 dimers) and three
  non-trivial ones (formed by 4 dimers).}
\label{fig:fig2}
\end{figure}

Consider now the scalar product of two such functions $ \Psi_1$, $ \Psi_2$: 
\be
S_{ij} = \brak \Psi_1 | \Psi_2 \kket
\ee
Let us draw both functions on a lattice. Such a situation can be viewed as a
set of {\em closed polygons}. Every such a polygon is formed by dimers
belonging to $\Psi_1 , \Psi_2, \Psi_1 , \Psi_2, \dots$ (in alternating
manner);  the number of bonds 
forming this polygon is {\em even}. It is illustrated on the 
 Fig.~\ref{fig:fig2}.

Consider first the situation where  wave functions $\Psi_1 , \Psi_2$ 
correspond to {\em single} non-trivial polygon on the lattice. (We call the
polygon nontrivial if it is not a 'double bond', i.e. if its
length is  $2L$, $L>1$). It is a matter of straighforward calculation (it
follows also from the 'reduction principle', see below) to show that
\cite{LiDoAn88} 
\be
\brak \Psi_1 | \Psi_2 \kket = \frac{1}{2^{L-1}}.
\label{1Wielobok}
\ee
This result can be generalized to the situation where $\Psi_1, \Psi_2$
correspond to family of  polygons: $P_1$ of length
 $2L_1$, $\dots$, $P_k$ of length $2L_k$. In such a case, we have
 \cite{LiDoAn88}:  
\be
\displaystyle
\brak \Psi_1 | \Psi_2 \kket = 2^{- [\sum_{i=1}^k (L_i - 1)] }.
\label{WieleWielobk}
\ee
\section{Matrix elements of exponens of the Heisenberg operator: two-spin form}
\label{sec:h2}
In this Section, we consider the model with nearest-neighbour
interactions. (Although the main interest of this paper are frustrated models,
considerations of this section can be treated as a warm-up and a presentation
of techniques used in the next section). The Hamiltonian is of
the form
\be
H=J\sum_{\brak ij \kket} h_{ij},
\label{HamHeis_nn}
\ee
where $h_{ij} = {\bf s}_i\cdot {\bf s}_j$.
 
Write an expression (\ref{HamHeis_nn}) in the form
\be
H=H_1+H_2 + H_3 + H_4,
\label{rozbicie_h2}
\ee
in such a way that $H_1, \dots, H_4$ are sums of {\em commuting}
operators. One possible way to achieve this goal is as follows. Every site
index $i$ is in fact a two-index:  $i=(i_x, i_y)$, where $i_x$ is horizontal
index on the lattice and  $i_y$ is vertical index. Divide the Hamiltonian
(\ref{HamHeis_nn}) into $H_1, \dots, H_4$ in the following way:

\bi
\item $H_1$ is a sum of these operators $h_{ij}$ where $i,j$ are of the form:
  $(2k,2l),(2k,2l+1)$. Denote it as type 1;
\item
 $H_2$ is a sum of $h_{ij}$'s where  $i,j$ are of the form: $(2k,2l),(2k+1,2l)$ -- type 2;
\item
$H_3$ is a sum  of $h_{ij}'s$ where $i,j$ are of the form: $(2k,2l+1),(2k+1,2l+1)$ --
type 3;
\item
 $H_4$ is a sum of $h_{ij}$'s where $i,j$ are of the form: $(2k+l,2l),(2k+1,2l+1)$
-- type 4.
\ei

Consider now a matrix element $\brak \Psi_I |\exp(K H_k) | \Psi_J \kket$,
$k=1,\dots,4$ in order to check its positivity. Every operator $H_k$ is a sum of {\em commuting}
operators, so, if  $\disty H_k = \mathop{\sum}_{i,j {\rm\; of\; type}\; k} h_{ij}$, then
$\disty \exp (K H_k) = \mathop{\prod}_{i,j\; {\rm of\; type} \;k}\exp(K h_{ij})$ --
so one can write:
\[
\brak \Psi_I |\exp(K H_k) | \Psi_J \kket = 
\]
\[=\sum_{\al_1}
\sum_{\al_2} \dots \sum_{\al_M}
 \brak \Psi_I |\exp(Kh_{i_1 j_1})|\Psi_{\al_1} \kket
\brak \Psi_{\al_1} |\exp(Kh_{i_2 j_2})|\Psi_{\al_2} \kket
\dots 
\brak \Psi_{\al_M} |\exp(Kh_{i_M j_M})\Psi_{J} \kket
\]
 (here $M$ is a number of operators $h_{ij}$ in $H_k$). 
We can make a conclusion that {\em if matrix elements of the operator
  $\exp(Kh_{ij})$ is positive, then  the matrix element of $\exp(K H_k)$
  is also positive. } 
 
Let us calculate the matrix element of two-spin operator $h_{ij}$:
\be
\brak \Psi_1 |\exp( K h_{ij}) | \Psi_2 \kket
\ee
where $| \Psi_1\kket,  |\Psi_2\kket$ are dimer functions,
  $K=-\beta J/N$. Notice that antiferromagnetic case $J>0$ imply
 $K\in ]-\infty, 0[$.

Below, the following notation will be useful: 
\[
\eps^1=\exp(K\slash 2),\;\;\;
\eps_3= \exp(-3K\slash 2).
\] 
Notice that $\eps^1\in ]0,1[$, $\eps_3\in ]1,\infty[$.

Let us consider first the situation where $i,j$ are nearest neighbours.

\begin{figure}
\includegraphics{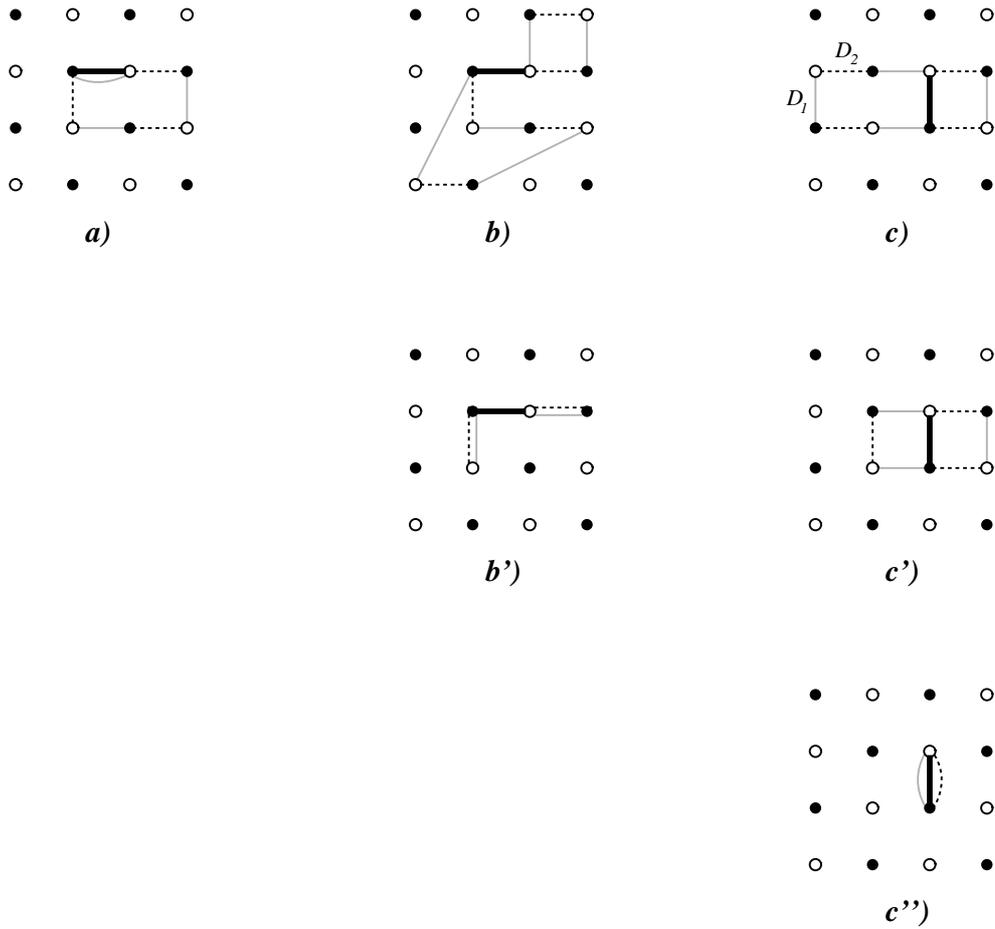}
\caption{{\bf a)}, {\bf b)}, {\bf c)} -- three kinds of matrix elements of the
  operator $\exp(Kh_{ij})$ listed above. The
  action of this operator is denoted as a bold line; gray and dashed lines
  denote dimers 
  forming the functions $\Psi_1$ and $\Psi_2$, respectively. {\bf c')} is
  obtained from  {\bf c)} by one step of reduction principle {\bf R.I}; upon
  this operation, neighbouring dimers $D_1$ and $D_2$ are eliminated.  {\bf
  b')} is obtained from {\bf b)} by three steps of reduction  {\bf R.I}. {\bf
  c'')} has been obtained from  {\bf c')} by two steps of  {\bf R.II}.   }
\label{fig:fig3}
\end{figure}

We have three sorts of situation:

{\bf a)}  The operator $\exp(Kh_{ij})$ acts on
  $|\Psi_1\kket$, which contains the $(ij)$ dimer; it is illustrated on 
 Fig.~\ref{fig:fig3}a.
 The value of the    matrix element is
\be
\brak \Psi_2|\exp(Kh_{ij}) |\Psi_1 \kket = 
\eps_3 \brak \Psi_2 |\Psi_1 \kket;
\label{ElMac_h2_Sytuacja1}
\ee
we see that this element is {\em positive.}

{\bf b)} The operator $\exp(Kh_{ij})$ is localized on a bond connecting two
different polygons; see  Fig.~\ref{fig:fig3}b.
 We have:
\be
\brak \Psi_2|\exp(Kh_{ij}) |\Psi_1 \kket
= \frac{1}{4}(3\eps^1+\eps_3) \brak \Psi_2 |\Psi_1 \kket
\label{ElMac_h2_Sytuacja2}
\ee
which is also {\em positive.}

{\bf c)} The operator $\exp(Kh_{ij})$ acts inside one connected polygon, but
  there are at   least three polygon 'bonds' between $i$ and $j$ sites; see
  Fig.~\ref{fig:fig3}c. 
 In such a situation we  have:
\be
\brak \Psi_2|\exp(Kh_{ij}) |\Psi_1 \kket 
=  \eps_3  \brak \Psi_2|\Psi_1 
\kket;
\label{ElMac_h2_Sytuacja3}
\ee
clearly, it is also {\em positive.}


Similar matrix elements (but of the operator $h_{ij}$ instead of its exponent)
have been calculated in \cite{LiDoAn88}. In the case considered here, they
have been derived by straightforward calculation with the
use of simple algebraic tools: spectral resolution of the operator $h_{ij}$
and the 'reduction principle'. Details of calculation of matrix elements is
presented below.

A matrix element for situation {\bf a)} (presented on Fig.~\ref{fig:fig3}a)
can be calculated 
immediately. Namely, in this situation the operator $\exp(Kh_{ij})$ acts for
dimer function  $(ij)$, which is  an {\em  eigenfunction} of $h_{ij}$ (and of
course of $\exp(Kh_{ij})$). Using property (\ref{fA_przezRzuty}) we get the
expression  (\ref{ElMac_h2_Sytuacja1}).

For situations {\bf b}  and {\bf c)}  (presented on Figs.~\ref{fig:fig3}b  and
  \ref{fig:fig3}c), we apply the {\em reduction principle} first. Such a
  reduction is performed in two steps {\bf R.I} and {\bf R.II}.

{\bf R.I} We examine the quotient  $q$:
\be
q=\frac{\brak \Psi_2|\exp(Kh_{ij}) |\Psi_1 \kket}{\brak \Psi_2|\Psi_1
  \kket}
\label{qNieRed}
\ee
 $\Psi_1$ and  $\Psi_2$ functions form closed (not necessarily
connected) polygon $P$. Assume that these functions can be written as:
$\Psi_1= \Phi_1 (lk)(nm)$, $\Psi_2=\Phi_2(lm)(np)$. Let us remove the
site $l$ and identify $k$ and $m$ sites (this situation can be viewed
as removing dimers $D_1=(kl)$ from
$\Psi_1$ and  $D_2=(mk)$ from $\Psi_2$ with subsequent
identifications of   sites $k$ and $m$). 
This way, we obtain 'reduced' functions
$\Psi_1^R= \Phi_1 (nk)$ and  $\Psi_2^R=\Phi_2 (np)$ and 'reduced'
 polygon $P^R$. We assume that all sites $k,l,m,n,p$ are at a distance
 greater than 1 from $i$ and $j$. 
Then, the quotient $q$ obtained for functions $\Psi_1$ and $\Psi_2$
 (\ref{qNieRed})  is
 equal to the quotient $q^R$ calculated for functions $\Psi_1^R$ and
 $\Psi_2^R$: 
 \be
q^R=\frac{\brak \Psi^R_2|\exp(Kh_{ij}) |\Psi^R_1 \kket}{\brak
 \Psi^R_2|\Psi^R_1
  \kket}
\label{qRed}
\ee
Less formally, we can say that the quotient $q$ will not change if we
remove two neighbouring dimers from the polygon $P$.

{\bf Proof.} Let us remember how to calculate the scalar product for
dimer wave functions: After expressing them in the 'plus-minus' basis,
we sum over all sites and spin degrees of freedom for every site. 

Let us consider the matrix element ${\brak \Psi_2|\exp(Kh_{ij})
  |\Psi_1 \kket}$. Let us expand dimer functions
$(lk),(nm),(lm),(np)$ in the base of 'plus-minus' functions. Then, let
us sum over spin indices of sites $l$ and $m$. After straightforward
calculations, remembering about normalization factors for dimers and
using orthonormality relation of spin functions on arbitrary site $r$:
$\brak r_\si | 
r_{\si'}\kket=\delta_{\si \si'}$, we obtain:
\be
{\brak \Psi_2|\exp(Kh_{ij}) |\Psi_1 \kket} = \frac{1}{2} {\brak
  \Psi^R_2|\exp(Kh_{ij}) |\Psi^R_1 \kket}
\ee
Analogous calculation gives
\be
{\brak \Psi_2|\Psi_1 \kket} = \frac{1}{2} {\brak
  \Psi^R_2|\Psi^R_1 \kket}
\ee
so we obtain desired equality $q=q_R$.

After (possibly, multiple) application of reduction principle to
situations as in Figs.~\ref{fig:fig3}b  and
  \ref{fig:fig3}c, we get situations such as on Figs.~\ref{fig:fig3}b'  and
  \ref{fig:fig3}c'. This way, reduction principle makes possible
  the calculation of matrix elements for situations, where wave
  functions occupy at most 6 sites.

{\bf R.II} By straightforward calculation one obtains that both cases
illustrated on Figs.~\ref{fig:fig3}c'  and~\ref{fig:fig3}c'' give equal
value of $q$. (This step is not necessary, 
  but its analogon for plaquettes will be useful  due to \oe{}conomy reasons).    

Two cases obtained after reduction, i.e. the ones illustrated on
  Figs.~\ref{fig:fig3}b'  and \ref{fig:fig3}c' (or, equivalently,
  \ref{fig:fig3}c'')  can be calculated immediately. The line of calculations
  is as follows: For the wave function  $\Psi_1$, one passes from dimer form
  to the  'plus-minus' basis. Then, one calculates an action of the operator
  $\exp(Kh_{ij})$ on $\Psi_1$,  making use of spectral resolution of the
  operator $h_{ij}$  given by (\ref{Ham_h2}), 
(\ref{Projectors_h2}) as well as formula
(\ref{fA_przezRzuty}). And last, one calculates the scalar product of
  expression obtained with the  $\Psi_2$ function, expressed in the
  'plus-minus' basis.

We can conclude this section by an assertion
 that  {\em matrix elements for nearest neighbour Heisenberg model
are positive, so we have no minus sign problem  here}.
 In my opinion, such a result can be viewed as an interesting one, but not
 exciting:  There are other approaches, where
minus-sign problem has been overcomed \cite{Manousakis}, \cite{Hirsch},
\cite{Sandvik}. 

Now, let us consider the frustrated case (i.e. the $J_1-J_2$ Heisenberg
model). In this case, however, the matrix elements are in general {\em not
  positive}. As an example, let us mention situation analogous to the case
{\bf b)}  above, but where $i,j$ are next-nearest neighbours. In such
a case, we have 
\be
\brak \Psi_2|\exp(Kh_{ij}) |\Psi_1 \kket 
=- \frac{1}{4}( 3\eps^1 + \eps_3)  \brak \Psi_2|\Psi_1 
\kket, 
\label{Psi2ExpKhijPsi1_2}
\ee
which is {\em not  positive}. 
In the other words: For {\em frustrated} ($J_1-J_2$) model, where both nn and nnn
  interaction are present, the minus-sign problem still exists.

 How to cure the problem? The idea is as follows: Write the Hamiltonian as a sum
  over {\em plaquettes} (4-spin) sets. One can hope that negative
  contribution will be compensated by positive one. (It is not obvious {\em a
  priori}, as two-body operators entering into plaquette operator 
  does not commute in general). It turns out that {\em in such a
  formulation, the matrix 
  elements are positive;} details are described in the following Section.

\section{Matrix elements of exponens of the Heisenberg operator: plaquette
  form}
\label{sec:h4}


Consider the Heisenberg model on a (subset of) square lattice. We assume that
there are  nn and  nnn interaction. For concreteness, we consider the
$J_1-J_2$ model, but all considerations apply also in the case of
'pyrochlore' lattice and some others. We assume that 
the system exhibits the reflection symmetry (remember in such a case, the
ground state is singlet). 

The setup for wave functions is the same as previously: We assume that wave
functions are built up from 'bipartite' dimers. 

\begin{figure}
\includegraphics{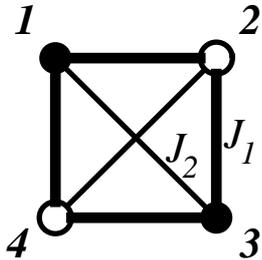}
\caption{Elementary plaquette on which the plaquette hamiltonian
  (\ref{HamPlak}) is defined. Sites are arranged anticlockwise. }
\label{fig:fig4}
\end{figure}

Consider the Hamiltonian defined on a square plaquette:
\be
h_\plak = h_{12} + h_{23} + h_{34} + h_{41} + \al(h_{13} + h_{24}),
\label{HamPlak}
\ee
(see Fig.\ref{fig:fig4}), 
where: $h_{ij}= {\bf s}_i \cdot {\bf s}_j$,  $\al = J_2\slash{}J_1$.

We have:
\be
 H = \sum_{i\in\La} h_{\plak,i}
\label{HamJakoSumaOfHamPlak}
\ee

Analogously as before, let us write the Hamiltonian in the form

\be
H=H_1+H_2 + H_3 + H_4,
\label{rozbicie_h4}
\ee
where each of terms $H_1,\dots, H_4$ is a sum of commuting plaquette
operators.
This division can be done, for instance, in the following form. Plaquette
index in (\ref{HamJakoSumaOfHamPlak}) is in fact a two-index:
 $i=(i_x,i_y)$, where $i_x$ --  horizontal component, 
 $i_y$ -- vertical one.  $H_1, \dots, H_4$ are defined as:
\bi
\item $H_1$ is a sum of these operators  $h_{\plak,i}$ where $i$ is of the
  form: $(2k,2l)$. Denote it as  type 1;
\item $H_2$ is a sum of these operators  $h_{\plak,i}$ where $i$ is of the
  form:  $(2k+1,2l)$. Denote it as  type 2;
\item $H_3$ is a sum of these operators  $h_{\plak,i}$ where $i$ is of the
  form: $(2k,2l+1)$. Denote it as  type 3;
\item $H_4$ is a sum of these operators  $h_{\plak,i}$ where $i$ is of the
  form: $(2k+1,2l+1)$. Denote it as  type 4.
\ei

As in the previous Section one shows that if matrix elements of elementary
 plaquette operator
 $\brak \Psi_I |\exp(K h_\plak)
| \Psi_J \kket$ are positive, then  matrix elements $\brak \Psi_I |\exp(K
H_k) | \Psi_J \kket$ are also positive.

Let us calculate the matrix element of $h_\plak$. This will be done in two
stages. In the first step we apply the 'reduction principle'. Its first stage
{\bf R.I.plaq} is identical as in the case of two-site hamiltonian {\bf R.I},
i.e. it allows to decrease the number of neighbouring polygon edges by
two without change 
of the quotient  $q_\plak$, where $q_\plak$  is
\be
q_\plak=\frac{\brak \Psi_2|\exp(Kh_\plak) |\Psi_1 \kket}{\brak \Psi_2|\Psi_1 \kket}
\label{qPlak}
\ee
The proof of {\bf R.I.plaq} is also identical as
previously, so we will not repeat it.  This way, all  calculations are reduced
to cases where wave functions occupy at most 12 sites. It is possible further
reduction. This second step of reduction is similar to {\bf R.II} but not
identical:

{\bf R.II.plaq:} Assume that in the configuration there appear a square
consisting of the following sides: One edge belong to the hamiltonian
plaquette (say, this is  $(i,j)$ side); two sides are dimers belonging to the
function $\Psi_1$: $(i,l)$, $(k,j)$; and the last side is dimer belonging to
the function  $\Psi_2$: (let it be $(k,l)$). In such a configuration, one can
eliminate two sides  $(k,j)$ and $(k,l)$ and replace the square $(i,j,k,l)$ by
one bond $(i,j)$ with dimer
$(i,j)$. 

\begin{figure}
\includegraphics{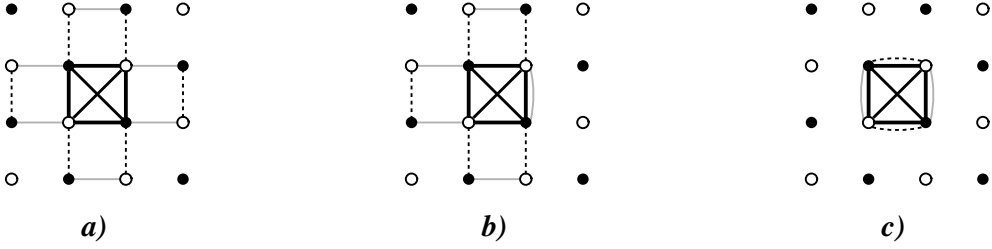}
\caption{Illustration of the second step reduction (\mbox{\bf R.II.plaq})
  action. a) corresponds to initial configuration; b) is the configuration
  after single action of (\mbox{\bf R.II.plaq}); on c), after three further actions
  of (\mbox{\bf R.II.plaq}). All three configurations have the same
  value of $q_\plak$.}  
\label{fig:fig5}
\end{figure}

The proof of {\bf R.II.plaq} can be  obtained by a straightforward
calculation. An example of its action  is illustrated on Fig.~\ref{fig:fig5}.   

This way, after reductions, we have to  calculate matrix elements of the
plaquette hamiltonian  with wave functions defined on at most 8 sites. 

\begin{figure}
\includegraphics{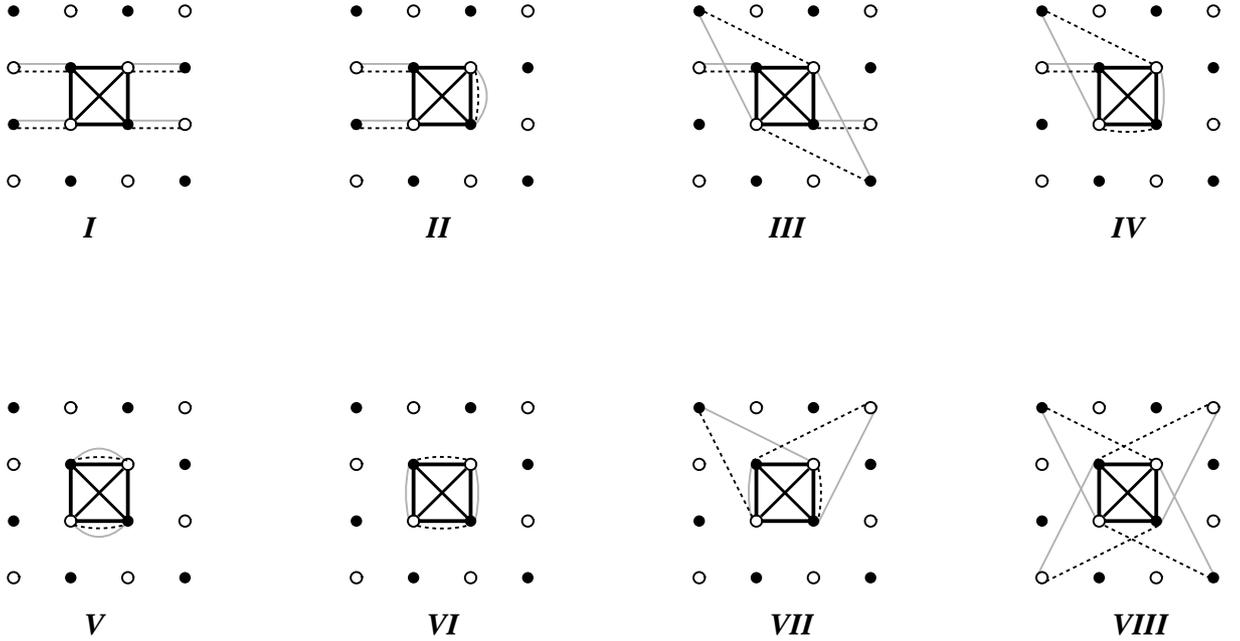}
\caption{Reduced configurations for plaquette Hamiltonian. }
\label{fig:fig6}
\end{figure}

It turns out that there are eight such configurations. They are illustrated on
 Fig.\ref{fig:fig6}.
 The matrix elements corresponding to them
 can be calculated in a straightforward  manner -- similarly as  in the previous
 Section for the case of two-spin Hamiltonian. Basic steps of such calculations are:
\ben
\item We express dimer functions in the 'plus-minus' basis.
\item We calculate the action of the $\exp(Kh_{\plak})$ operator on the
  function $\Psi_1$, using the spectral resolution of the operator $h_\plak$
  (necessary formulas are collected in Sec. \ref{subsecA:h4}) together
  with the formula   (\ref{fA_przezRzuty}). 
\item And last, we calculate the scalar product of the expression obtained
  above with the wave function  $\Psi_2$ (expressed in the 'plus-minus'
  basis). 
\een
 
Calculations are  straightforward  but lengthy, and they have been
 performed with the use of
  symbolic calculations programme (Maple).
 The results are  summarized in the Table below. 

In formulas below, the following notation has been used $y=\exp(-K)$;
$a=\alpha\slash{}2$. We assume that  $0\leq \al\leq 1$
(as for these values of $\al$ we have warranty that the ground state of the
system is singlet \cite{S3}, \cite{LSW}). Remember that for  $0\leq T \leq
\infty$, we have  $1\leq y \leq\infty$. 

\pagebreak

{\bf Table:} {\em Values of matrix elements of the quotient $q_{\plak}$ for eight
reduced  configurations on Fig.~\ref{fig:fig6}}.
\vskip.1cm
\centerline{
\begin{tabular}{|c|c|}
\hline \hline
No & $q_\plak$\\
\hline
  I & $\frac{1}{16}\left(  y^{3a} + y^{2-a} + 6 y^a + 3 y^{1-a} + 5
    y^{-1-a} \right)$\\
\hline
  II & $\frac{1}{16}\left(  y^{3a} + 3 y^{2-a} + 6 y^a + 6 y^{1-a} 
     \right)$\\
\hline
  III & $\frac{1}{8}\left( - y^{3a} + y^{2-a} + 3 y^{1-a} + 5 y^{-1-a} 
     \right)$\\
\hline
  IV & $\frac{1}{8}\left( - y^{3a} + 3 y^{2-a}  + 6 y^{1-a} 
     \right)$\\
\hline
  V & $\frac{1}{4}\left(  y^{3a} + 3y^{2-a}  \right)$\\
\hline
  VI & $\frac{1}{2}\left( - y^{3a} + 3 y^{2-a} \right)$\\
\hline
  VII & $\frac{1}{4}\left(  y^{3a} + 3 y^{2-a} -6 y^a + 6 y^{1-a}  
    \right)$\\
\hline
  VIII & $\frac{1}{4}\left(  y^{3a} + y^{2-a} -6  y^a +3 y^{1-a} + 
   5 y^{-1-a} \right)$\\
\hline\hline
\end{tabular}
}
\vskip.2cm
Some of expressions in the Table are evidently positive (there are:  I, II,
V). It turns out that remaining expressions are also positive; proofs are
presented below. Remember that $a\in [0,\frac{1}{2}]$ and 
$y\in]1,\infty[$. Factors before expressions for $q_\plak$  are skipped.
\bi
\item {\bf Situation III.} We have:
\[
 - y^{3a} + y^{2-a} + 3 y^{1-a} + 5 y^{-1-a} \geq  - y^{3a} + y^{2-a}= y^{3a}(-1+y^{2-4a});
\]
but for $2-4a\geq 0$ and for $y>1$ we have $(-1+y^{2-4a})>0$.
 \item {\bf Situation IV.}
\[
 - y^{3a} + 3 y^{2-a}  + 6 y^{1-a} \geq  - y^{3a} + y^{2-a},
\]
and  this is the same expression as in III. 
 \item {\bf Situation VI.} 
\[
 - y^{3a} + 3 y^{2-a} \geq  - y^{3a} + y^{2-a},
\] 
so we again obtain case studied in III.
\item {\bf Situation VII.}
\[
 y^{3a} + 3 y^{2-a} -6 y^a + 6 y^{1-a} \geq 6(-y^a+y^{1-a}) = 6 y^a(-1+y^{1-2a})
\] 
and, analogously as in III, conditions: $1-2a\geq 0$, $y>1$ imply that
$-1+y^{1-2a}>0$. 
\item {\bf Situation VIII.} This is the most complicated one. Write first:
\[
 y^{3a} + y^{2-a} -6  y^a +3 y^{1-a} +   5 y^{-1-a} =  y^{3a} + y^{2-a} -3 y^a
 +   5 y^{-1-a}+3(-  y^a+ y^{1-a}).
\]
Due to argumentation identical as in III, the last term (in parentheses) is
 positive:
 \mbox{$ -
y^a+ y^{1-a}= y^a(-1+y^{1-2a}) \geq 0$}. So, it is sufficient  to show positivity of
\mbox{$ y^{3a} + y^{2-a} -3 y^a +   5 y^{-1-a}= y^a(y^{2a} +
y^{2-2a}-3+5y^{-1-2a})$}. 

Consider two extreme cases (i.e. $a=0$ and $a=\frac{1}{2}$) of the expression
in parentheses above. We have for $a=0$:
\[
1+y^2-3+5y^{-1}> y-2+\frac{1}{y} + \frac{4}{y} =
\left( \sqrt{y} - \frac{1}{\sqrt{y}}\right)^2 + \frac{4}{y} >0,
\] 
and for $a=\frac{1}{2}$
\[
y+y-3+5 y^{-2}=y^{-2}(2y^3-3y^2+5)=y^{-2}(y+1)(2y^2-5y+5),
\]
which is greater than zero for $y\geq 1$, as it follows from elementary
considerations.

For intermediate values of $a$, i.e. $0<a<\frac{1}{2}$, let us write
\[
y^{2a} + y^{2-2a}-3+5y^{-1-2a}= y^{-1-2a}(y^{1+4a}+y^3-3y^{1+2a}+5)
\]
and notice that $y^{1+4a}>y$, $-y^{1+2a}>-y^2$, so we have
\[
y^{1+4a}+y^3-3y^{1+2a}+5>y+y^3-3y^2+5=(y+1)(y^2-4y+5)>0,
\]
thus establishing positivity of VIII.
\ei

Let us summarize this section by the statement that {\em in the dimer basis,
  and for the Hamiltonian written as a sum of plaquettes, matrix elements of
  the $\exp(Kh_\plak)$ are positive}.

\section{Summary, conclusions}
\label{sec:Conclus}

The technical tool to study (certain class of)
 frustrated systems has been developed,
which (hopefully) would  allow application of probabilistic techniques. 

This paper was devoted to the  $J_1-J_2$ Heisenberg model in any dimension
(and models whose 
Hamiltonians can are sums of frustrated plaquette hamiltonians; the pyrochlore
lattice is perhaps the most typical example). It would be tempting to
generalize the method to other frustrated systems. Such a generalization
probably can be realized in the case of  other systems exhibiting reflection
symmetry, for instance 
3d Heisenberg models with frustrated cubes; of course, one have to calculate
the matrix elements of Hamiltonians for frustrated units. 

Generalization for frustrated systems exhibiting no reflection symmetry
(such as kagom\'{e} or triangular ones in two dimensions) is less obvious. For
such systems, matrix elements of frustrated units can be positive (author
calculated such elements for triangular lattice and positivity holds also in this
case). But, on the other hand, the method relies heavily on the assumption
that we are working in the singlet sector.
It has been proved that the ground state(s)  of the $J_1-J_2$ Heisenberg model is
singlet \cite{LS1}--\cite{S3}, \cite{LSW}, but for triangular or kagom\'{e}
lattices it is not known. (The answer is probably positive, but the proof --
as far I know -- is lacking). If the ground state is singlet, then also in the
case of triangular lattice one can try to simulate the ground state using this
method. 

The next point is the technical one: Full Monte Carlo simulation of dimer-like
models is rather difficult task \cite{Sandvik03}. It is not clear how
difficult would be an implementation of the actual method; this paper is
devoted only to development of the scheme. However, certain
attempts towards concrete computational realization of this method are in progress.

It would be interesting to try to develop analogous method in the sector of
$S$ arbitrary, not only $S=0$. If successful, it would be possible to perform
qMC simulations in arbitrary temperatures, not only in low-T region (as in the
present version). Moreover, it would be also possible to simulate systems
where is no warranty that the ground state is singlet (for instance, the
triangular lattice). There are some indications that the procedure described
in this paper
could be generalized for  $S$ arbitrary. However, at this moment it is too
early to say something definite.

\section{Appendix: Techniques used in calculation of matrix elements}
Let  $A$ be the self-adjoint operator in finite-dimensional Hilbert space
$\Hcal$, dim$\,\Hcal = N$; let  Sp$(A) = \{  \la_i \}$ --  spectrum of $A$, $V_i$ --
the subspace corresponding to the eigenvalue   $\la_i$, dim$\,V_i=n_i$. We
have: $\sum_i n_i = N$,  $\disty \Hcal
= \mathop{\oplus}_i V_i$ . Every such operator $A$ can be represented in the
form of the {\em spectral resolution}:
\be
A=\sum_i\la_i P_i,
\label{RozklSpektr}
\ee
where  $P_i$ -- orthogonal projection onto corresponding subspace $V_i$. These
projections possess well known properties:  $\sum_i P_i =Id_\Hcal$ ($Id_\Hcal$ -- the
identity operator in $\Hcal$);
$P_i^2=P_i$ for every $i$; $P_i P_j = P_j P_i = 0$ for $i \ne j$. Every such
projection $P_i$ onto eigensubspace $V_i$ can be calculated from the famous
formula 
\be
P_i=\sum_{k=1}^{n_i} | v_{i_k}\kket \brak v_{i_k}|,
\label{WzorNaRzut}
\ee
where  $|v_{i_k}\kket$ is  $k$-th vector of orthonormal base spannning the
$V_i$ subspace. 

If $f(x)$ -- analytic function, then we have: 
\be
f(A) =\sum_i f(\la_i) P_i
\label{fA_przezRzuty}
\ee

For the sake of completeness, we present below expressions for projections,
which appear in the spectral resolutions of operators $h_{ij}$ and $h_\plak$. 

\subsection{Spectral resolution of the operator $h_{ij}$}
\label{subsecA:h2}
For two spins, the state space 
 $\Hcal_2$ is four-dimensional. The Hamiltonian 
 $h_{ij} = {\bf s}_i\cdot {\bf s}_j$ commutes with the total spin
 operator, so eigenvalues of $h_{ij}$ can be classified accordingly to angular
 momentum quantum numbers. Eigenvalues of $h_{ij}$ are: $E_0=-\frac{3}{4}$
(total spin $S$ is zero, i.e. the state is singlet; its multiplicity is one),
 and $E_1=\frac{1}{4}$ 
(here total spin $S$ is one, i.e. it is a triplet. We have three
 triplet states with $z$-th component of angular momentum equal to
 $+1,0,-1$; all of them have the same energy, so
 the  multiplicity  of $E_1$ is 3). 

Let us choose in  $\Hcal_2$ the standard  ('plus-minus') basis $e_1, \dots, e_4$:
\[
e_1=|i_+j_+\kket,e_2=|i_+j_-\kket,e_3=|i_-j_+\kket,e_4=|i_-j_-\kket,
\]
In this basis, the Hamiltonian $h_{ij}$ is given by the matrix:
\be
h_{ij} = {\bf s}_i\cdot {\bf s}_j
=
\frac{1}{4}
\left[
\begin{array}{cccc}
1&0&0&0\\
0&-1&2&0\\
0&2&-1&0\\
0&0&0&1
\end{array}
\right],
\label{Ham_h2}
\ee
Projectors  $P_0$ (onto singlet subspace) and $P_1$ (onto triplet subspace) are:
\be
P_0=
\frac{1}{2}
\left[
\begin{array}{cccc}
0&0&0&0\\
0&1&-1&0\\
0&-1&1&0\\
0&0&0&0
\end{array}
\right],
\;\;\;
P_1=
\frac{1}{2}
\left[
\begin{array}{cccc}
2&0&0&0\\
0&1&1&0\\
0&1&1&0\\
0&0&0&2
\end{array}
\right].
\label{Projectors_h2}
\ee

\subsection{Spectral resolution of the operator  $h_\plak$}
\label{subsecA:h4}
The space of states for the system of four spins 
 $\Hcal_4$ is $2^4= 16$-dimensional. In such a situation, it is again very
 useful to 
 exploit properties of the angular momentum operator and classify states
 accordingly to spin quantum numbers.

 The plaquette Hamiltonian is given by (\ref{HamPlak}). Remember $\al =
 J_2\slash{}J_1$; it is more convenient to introduce $a=\al\slash{}2$. The
 Hamiltonian (\ref{HamPlak}) commutes with total angular momentum
 operator $ {\bf
 S }$. For eigenstates and eigenvalues of the Hamiltonian, good quantum numbers
 are  $S$ (total spin) and  $M$ ($z$-th component of spin). Let us classify states
 accordingly to the value of  $M$ first, and then, in every sector with given
 $M$, classify the states accordingly to $S$. For each such state, we will
 give their energies and corresponding projectors. Classification of states
 with respect to values of $M$ proceeds as follows.

Among all 16 states, we have:
\bi
\item One state with $M=2$  (plus twin state $M=-2$). Both of them share $S=2$.
\item Four states with $M=1$. Among them, there are: One state $S=2,M=1$ and
  three states $S=1,M=1$. There are also twin states for $M=-1$.
\item six $M=0$ states. There are: one state $S=2,M=0$; three states
  $S=1,M=0$; two states $S=0,M=0$.
\ei
We can consider only states with non-negative values of $M$, as all of them
  possess their twins for $-M$.

Consider now all $M$ sectors. Eigenvalues will be denoted as :
$E^{(n)}_{S,M}$ ($n$ -- index of the state), and corresponding projections by
$P^{(n)}_{S,M}$.

\bi
\item
{\bf  $M=2$ sector}.

This is subspace spanned by one base vector 
$e^{(2)}_1=\pppp$. The Hamiltonian in this sector is simply a number $1+a$,
which is of course also the eigenvalue  $E^{(1)}_{2,2}$. The projection is
trivial.

\item {\bf  $M=1$ sector}.

This subspace is four-dimensional. As a basis, let us choose:
\be
e^{(1)}_1=\mppp, e^{(1)}_2=\pmpp, e^{(1)}_3=\ppmp, e^{(1)}_4=\pppm
\label{m1}
\ee
The Hamiltonian in this basis is given by:
\be
h_\plak=
\pol\left[
\begin{array}{cccc}
0&1&2a&1\\
1&0& 1&2a\\
2a& 1&0&1\\
1&2a&1&0
\end{array}
\right]
\label{HamPlak_M1}
\ee 
Eigenvalues (i.e. energies) and corresponding projectors are:

\[
E^{(1)}_{2,1} = 1+a, \;\;\; \;\;
P^{(1)}_{2,1}=
\frac{1}{4}
\left[
\begin{array}{cccc}
1 &1 &1 & 1\\
1 &1 & 1&1 \\
1 & 1&1 & 1\\
 1&1 &1 &1
\end{array}
\right]
\]
(this is $M=1$ component of quintet). The remaining three states are triplets
(more precisely, the  $M=1$ components thereof). Two of them are degenerate:
the subspace spanned by eigenvectors 
of $E^{(1)}_{1,1}$ is two-dimensional:
\[
E^{(1)}_{1,1} = -a, \;
P^{(1)}_{1,1}=
\frac{1}{2}
\left[
\begin{array}{cccc}
1 &0 &-1 & 0\\
0 &1 & 0&-1 \\
-1 & 0&1 & 0\\
 0&-1 &0 &1
\end{array}
\right],
\]
\[
E^{(2)}_{1,1} = -1+a, \;
P^{(2)}_{1,1}=
\frac{1}{4}
\left[
\begin{array}{cccc}
1 &-1 &1 & -1\\
-1 &1 & -1&1 \\
1 & -1&1 &- 1\\
- 1&1 &-1 &1
\end{array}
\right]
\]

And last, consider 

\item {\bf The $M=0$ sector.}
The basis is:
\[
e^{(0)}_1= \ppmm, e^{(0)}_2=\mmpp, e^{(0)}_3= \pmmp, e^{(0)}_4=\mppm,
\]
\be
e^{(0)}_5=\pmpm, e^{(0)}_6=\mpmp 
\label{m0}
\ee
The Hamiltonian in $M=0$ sector is:
\be
h
=
\frac{1}{2}
\left[
\begin{array}{cccccc}
 -2a&0 &2a &2a &1 &1 \\
0 &-2a &2a &2a &1 &1\\
2a &2a &-2a &0 &1 &1\\
2a &2a &0 &-2a &1 &1\\
1 &1 &1 &1 &-2+2a &0\\
1 & 1& 1& 1&0 &-2+2a\\
\end{array}
\right]
\label{HamM0}
\ee 
Eigenvalues and projectors are:

-) The $M=0$ component of quintet:
\be
E^{(1)}_{2,0}=1+a,\;\;
P^{(1)}_{2,0}
=
\frac{1}{6}
\left[
\begin{array}{cccccc}
 1 & 1 & 1 & 1 & 1 & 1\\
 1 & 1 & 1 & 1 & 1 &1\\
 1 & 1 & 1 & 1 & 1 &1\\
 1 & 1 & 1 & 1 & 1 &1\\
 1 & 1 & 1 & 1 & 1 &1\\
 1 & 1 & 1 & 1 & 1 &1\\
\end{array}
\right]
\label{kwintetPlak}
\ee 
-) The $M=0$ components of triplets:
\be
E^{(1)}_{1,0}=-a,\;\;
P^{(1)}_{1,0}
=
\frac{1}{2}
\left[
\begin{array}{cccccc}
1 & -1 &0 & 0&0 & 0\\
 -1& 1& 0& 0& 0&0\\
0 &0 &1 &-1 &0 &0\\
 0& 0&-1 &1 &0 &0\\
0 &0 &0 &0 &0 &0\\
0 & 0&0 & 0&0 &0\\
\end{array}
\right]
\label{trypletPlak1}
\ee 
(remember that subspace spanned by eigenvectors corresponding to
$E^{(1)}_{1,0}$ is two-dimensional), and
\be
E^{(2)}_{1,0}=-1+a,\;\;
P^{(2)}_{1,0}
=
\frac{1}{2}
\left[
\begin{array}{cccccc}
 0 & 0 & 0 & 0 & 0 & 0\\
 0 & 0 & 0 & 0 & 0 &0\\
 0 & 0 & 0 & 0 & 0 &0\\
 0 & 0 & 0 & 0 & 0 &0\\
 0 & 0 & 0 & 0 & 1 &-1\\
 0 & 0 & 0 & 0 & -1 &1\\
\end{array}
\right]
\label{trypletPlak2}
\ee 
-) And last, singlets:
\be
E^{(1)}_{0,0}=-3a,\;\;
P^{(1)}_{0,0}
=
\frac{1}{4}
\left[
\begin{array}{cccccc}
 1 & 1 & -1 & -1 & 0 & 0\\
  1& 1 & -1 & -1 & 0 &0\\
 -1 & -1 & 1 & 1 & 0 &0\\
 -1 & -1 & 1 & 1 & 0 &0\\
 0 & 0 & 0 & 0 & 0 &0 \\
 0 & 0 & 0 & 0 &  0 &0\\
\end{array}
\right]
\label{singletPlak1}
\ee 
\be
E^{(2)}_{0,0}=-2+a,\;\;
P^{(2)}_{0,0}
=
\frac{1}{12}
\left[
\begin{array}{cccccc}
 1 & 1 & 1 & 1 & -2 & -2\\
  1& 1 & 1 & 1 & -2 &-2\\
 1 & 1 & 1 & 1 & -2 &-2\\
 1 & 1 & 1 & 1 & -2 &-2\\
 -2 & -2 & -2 & -2 & 4 &4 \\
 -2 & -2 & -2 & -2 &  4 &4\\
\end{array}
\right]
\label{singletPlak2}
\ee
 
\ei

\vskip.2cm
{\bf Acknowledgments.} I would like to express the
gratitude for  Prof. M. Fannes (KU Leuven) and
Dr. J. Tworzyd\l{}o (IFT UW)  for illuminating
discussions. This work was partly supported by the post-doctoral
Research  Training
Program HPRN-CT-2002-00277 and by the grant SPUB127 financed by
Komitet Bada\'{n} Naukowych, as well as  by the ESF  grant  Random
Dynamics in Spatially Extended Systems (visit in
Leuven).


\begin{thebibliography}{99}
\bibitem{WannHout}Wannier, G. H.: {\em Phys. Rev.} {\bf 79}, 357 (1950);
 Houtappel, R.M.F.: {\em Physica} {\bf 16}, 425 (1950).
\bibitem{FishSzp} Fisher,  M. E. and Szpilka,  A. M.: {\em Phys. Rev.}
 {\bf B  36}, 5343 (1987). 
 \bibitem{FishSel}Fisher, M. E. and Selke, W.: {\em Phys. Rev. Lett.}
{\bf 44}, 1502 (1980); {\em Phil. Trans. Roy. Soc. Lond.} {\bf A 302},
1 (1981).
 \bibitem{DinMaz}Dinaburg, E. I. and Mazel, A. E.:  {\em Comm. Math. Phys.}
{\bf 125}, 27 (1989).
  \bibitem{Selke} Selke, W. In: {\em Phase Transitions and Critical Phenomena},
(ed. by C. Domb and J. L. Lebowitz), vol. {\bf 15}, Academic Press, 1992.
\bibitem{FrIs} Liebmann, R.: {\em Statistical Mechanics of Periodic Frustrated
Spin Systems}. Springer, Berlin 1986. 
\bibitem{tt1HuM} Hlubina, R., Sorella, S. and  Guinea, F.: {\em
    Phys. Rev. Lett.} {\bf 78}, 1343 (1997); Avella, A., Mancini, F.,
    Villani, D. and Matsumoto, H.: {\em Eur. Phys. J.} {\bf B 20}, 303
    (2001);  Honerkamp, C. and Salmhofer, M.:
 {\em Phys. Rev.} {\bf B 64}, 184516 (2001).
\bibitem{Anderson73} Anderson, P. W.: {\em Mater. Res. Bull.} {\bf 8},
  153 (1973);  Fazekas, P. and Anderson, P. W.: {\em Philos. Mag.}
  {\bf 30}, 423 (1974).
\bibitem{Moessner} Moessner, R.: {\em Can. J. Phys.} {\bf 79}, 1283 (2001). 
\bibitem{FrSS} {\em Magnetic Systems with Competing Interactions: Frustrated
Spin Systems}. H.T. Diep (ed.), World Scientific, Singapore 1994.
\bibitem{AKLT} Affleck, I.,  Kennedy, T.,  Lieb, E. H. and  Tasaki, H.: 
{\em Phys. Rev. Lett.} {\bf 59}, 799 (1987);  {\em Comm. Math. Phys.}
{\bf 115}, 477 (1988). 
 \bibitem{ZittKlumpEtAl} Kl\"{u}mper, A.: {\em J. Phys. A: Math. Gen.}
 {\bf 23} , 809 (1990);  Kl\"{u}mper, A., Schadschneider, A. and  Zittartz,
 J.: {\em Z. Phys.} {\bf B 87}, 291 (1992); Niggemann, H., Kl\"{u}mper, A.
 and Zittartz, J.: {\em Z. Phys.} {\bf B 104}, 103 (1997).
\bibitem{SorellaEtAl} Capriotti, L.,  Becca, F.,  Parola, A. and  Sorella, S.: 
{\em Phys. Rev. Lett.} {\bf 87}, 097201 (2002).
\bibitem{LS1}  Lieb, E. H. and  Schupp, P.: {\em Phys. Rev. Lett.} {\bf 83},
  5362   (1999). 
\bibitem{LS2}   Lieb, E. H. and  Schupp, P.: {\em Physica} {\bf A 279},
  378  (2000).
\bibitem{S3} Schupp, P.: {\sf math-ph}\slash{}0206021.
\bibitem{Suzuki} Suzuki, M.: {\em Prog. Theor. Phys.} {\bf 56}, 1454 (1976); 
{\em Phys. Rev.} {\bf B 31}, 2957 (1985).
\bibitem{Kennedy85} Kennedy, T.: {\em Comm. Math. Phys.} {\bf 100}, 447 (1985).
\bibitem{ReedSimon1} Trotter, H. F.: {\em Proc. Amer. Math. Soc.} {\bf
  10}, 545
 (1958); Reed, M. and Simon, B.: Methods  of Modern Mathematical
 Physics, vol. 1 (Academic Press, 1972), Chap. 8. 
\bibitem{deRaedt_qMC} de Raedt, H. and von der Linden, W. In: {\em The
  Monte  Carlo Method in Condensed Matter Physics}, ed. by K. Binder,
  Topics in Applied Physics, vol. 71, Springer-Verlag, 1992.
\bibitem{AizenmanNachtergaele94} Aizenman, M. and Nachtergaele, B.: {\em
Comm. Math. Phys.} {\bf 164}, 17 (1994)
\bibitem{ALSSY2004} Aizenman, M.,  Lieb, E. H., Seiringer, R., Solovej,
  J. P. and Yngvason, J.:  {\em Phys. Rev.} {\bf A 70},
023612 (2004).
\bibitem{FeFroUel} Fernandez, R., Fr\"{o}hlich, J. and Ueltschi, D.: {\sf
    math-phys\slash{}0509060}; {\em Comm. Math. Phys.} {\bf 266}, 777 (2006). 
 \bibitem{Manousakis} Manousakis, E.: {\em Rev. Mod. Phys.} {\bf 63}, 1
  (1991).
\bibitem{Hirsch} Hirsch, J.: {\em Phys. Rev.} {\bf B 31}, 4403 (1985);
  {\em Phys. Rev.} {\bf B 35}, 1851 (1987).
\bibitem{Sandvik} Henelius, P. and  Sandvik, A. W.:  {\em Phys. Rev.}
  {\bf B 62}, 1102 (2000).
\bibitem{LiDoAn88} Liang, S., Doucot, B. and Anderson, P. W.: {\em
    Phys. Rev. Lett.}  {\bf 61}, 365 (1988).
 \bibitem{Caspers} Iske, P. L. and Caspers, W. J.: {\em Physica} {\bf A
 142}, 360 (1987). 
 \bibitem{Sutherland} Sutherland, B.:  {\em Phys. Rev.} {\bf B 37},
 3786 (1987).
\bibitem{LouSandvik06} Lou, J. and Sandvik, A. W.: {\sf cond-mat\slash{}0605034}.
\bibitem{LSW} Wojtkiewicz, J.: {\em Eur.  Phys. J.} {\bf B 44}, 501 (2005).
\bibitem{Sandvik03} Sandvik, A. W. and Moessner, R.:  {\em Phys. Rev.} {\bf B 73}, 144504 (2006).
\end{thebibliography}
\end{document}